\documentclass[twocolumn,showpacs,pra]{revtex4}
\usepackage{graphicx}
\usepackage{subfigure}
\usepackage{bm}

\begin{document}

\title{Bragg spectroscopy with an accelerating Bose-Einstein condensate}
\author{R. Geursen}
\email{reece@physics.otago.ac.nz}
\author{N. R. Thomas}
\author{A. C. Wilson}
\affiliation{Department of Physics, University of Otago, P.O. Box 56, Dunedin, New Zealand}
\date{\today}

\begin{abstract}
We present the results of Bragg spectroscopy performed on an
accelerating Bose-Einstein condensate. The Bose condensate undergoes
circular micro-motion in a magnetic TOP trap and the effect of this
motion on the Bragg spectrum is analyzed.  A simple frequency modulation
model is used to interpret the observed complex structure, and broadening
effects are considered using numerical solutions to the Gross-Pitaevskii
equation. 
\end{abstract}
\pacs{03.75.Kk, 03.75.Nt, 32.80, 32.80.Cy}
\maketitle

\section{\label{Intro}Introduction}

The experimental realization of a dilute Bose-Einstein condensate (BEC), a
quantum-degenerate atomic Bose gas, has led to a wide range of experiments
in atom optics (see, for example, Ref.~\cite{review,ketterlereview}). The
macroscopically occupied ground state exhibits large-scale coherence,
which can be exploited in atom optical techniques. One such technique is
Bragg scattering, which has been applied to make precise measurements of
condensate momentum \cite{MITspec,Braggspec}, investigate excitations \cite
{excitationspec}, measure the phase coherence length \cite{phasecoherence},
excite phonons \cite{phonons}, and to implement four-wave mixing \cite{4wave}
and a Mach-Zehnder interferometer \cite{machzeh}. More recently, Blakie 
\emph{et al.} showed that the velocity sensitivity of Bragg scattering can
provide a means of detecting superfluid flow associated with condensates in
a vortex state \cite{blairbraggfromvortex}. In all this work, the initial
Bose condensates have been well described as wave-functions with a
stationary center of mass (COM), making them an ideal source for atom-optic
experiments using Bragg scattering.

In conventional Bragg spectroscopy the spectrum is given by the spectral
response function of the condensate \cite{Braggspec}, which is related to
the momentum distribution and under certain conditions is a good measure of
this distribution. In this work, we show that for an accelerating
condensate, the spectrum is no longer a direct representation of the momentum distribution. We investigate Bragg spectroscopy for the case of a Bose
condensate undergoing circular micro-motion in a Time-averaged Orbiting
Potential (TOP) trap \cite{micromotion,TOP}. Although the condensate
micro-motion is small in amplitude, we demonstrate that the COM acceleration can be large enough to dramatically alter the Bragg spectrum.
Our experimental results are compared to a theoretical spectrum for the
corresponding stationary condensate, illustrating that the effect of the
acceleration is to significantly broaden the Bragg spectrum and introduce
complex structure. We describe two simple models to explain the underlying
physics of the observed behavior, and present numerical simulations using
the one-dimensional Gross-Pitaevskii equation (GPE) which confirm our
interpretation.

\section{\label{background}Condensate micro-motion and Bragg Spectroscopy}

In this section we summarize aspects of micro-motion in a TOP trap that are relevant to its affect on Bragg spectroscopy, and
present two simple models for describing the modified Bragg scattering
process.

The instantaneous magnetic field for our TOP trap is, as given in Ref.~\cite{TOP},

\begin{equation}  \label{instanB}
\mathbf{B}=(B^{\prime}_{q}x+B_{T}\cos (\omega _{T}t))\widehat{\mathbf{i}}%
+(B^{\prime}_{q}y+B_{T}\sin (\omega _{T}t))\widehat{\mathbf{j}}%
-2B^{\prime}_{q}z\widehat{\mathbf{k}},
\end{equation}

\noindent where $B_{q}^{\prime }$ is the quadrupole field gradient along the
radial direction, $B_{T}$ is the magnitude of the rotating bias field, and $
\omega _{T}$ is the angular frequency of the the bias field rotation. Petrich \emph{et al. }\cite{TOP} showed that, for time scales longer than the rotation period, the atoms experience only the leading terms in the magnitude of the time averaged magnetic field so that

\begin{equation}  \label{avgB}
B_{avg}\simeq B_{T}+\frac{B^{\prime2}_{q}}{4B_{T}}(r^{2}+8z^{2}),
\end{equation}

\noindent where $r$ is the radial coordinate. For many experiments, this
time-average is an appropriate description, but for Bragg spectroscopy this
is not always the case \cite{1Doptlat}. The time-varying force on the condensate results
in COM oscillation or micro-motion (as pointed out by Petrich \emph{et al.} 
\cite{TOP}), to which Bragg spectroscopy is sensitive. The origin of this micro-motion can be seen by considering the instantaneous potential of the TOP trap

\begin{eqnarray}  \label{instanpot}
U(x,y,z,t)&=&\mu |\mathbf{B}|  \nonumber \\
&=&\mu B^{\prime}_{q}[r_{0}^{2}+x^{2}+y^{2}+4z^{2}  \nonumber \\
& & +2r_{0}(x\cos (\omega _{T}t)+y\sin (\omega _{T}t))]^{\frac{1}{2}},
\end{eqnarray}

\noindent where $r_{0}=B_{T}/B_{q}^{\prime }$ is the so-called
\mbox{\textquotedblleft circle of death\textquotedblright\ } radius (the radius
where the \mbox{$B=0$} point rotates in the \mbox{$z=0$} plane) and the time-varying force
is given by \mbox{$\mathbf{F}(x,y,z,t)=-$$\nabla U$}. 
Typically, in the region of the condensate in the trap, \mbox{$r_{0}\gg x,y$} and we can approximate the magnitude of the force along the \emph{x}-axis (for example) as

\begin{equation}
F(t)\simeq -\mu B_{q}^{\prime }\cos (\omega _{T}t).  \label{approforce}
\end{equation}%
Ignoring condensate mean-field effects (since the trapping force is large)
we find that the classical solution for the magnitude of the time-varying COM momentum and position are given by

\begin{eqnarray} \label{oscilmomentum} 
p(t)& = &\frac{-\mu B^{\prime}_{q}}{\omega _{T}}\sin (\omega _{T}t) \\
\label{oscilposition}
x(t)& = &\frac{\mu B^{\prime}_{q}}{m\omega _{T}^{2}}\cos (\omega _{T}t),
\end{eqnarray}

\noindent where $m$ is the atomic mass and we have assumed for simplicity \mbox{$p(0)=0$}, and \mbox{$x(0)=\mu B^{\prime}_{q}/m\omega _{T}^{2}$}. The condensate COM along the \emph{x}-axis performs simple harmonic motion, and from trap symmetry we deduce that the condensate performs circular motion in the plane of the rotating bias field. This result was first obtained and experimentally confirmed by M\"{u}ller \emph{et al.} \cite{micromotion}.

To discuss the implications of centripetal acceleration for Bragg
spectroscopy we first briefly summarize Bragg scattering \cite{diffraction}.
A moving optical grating is applied to the condensate and atoms can absorb $
n $ photons from one beam ($\omega _{1},\mathbf{k}_{1}$) and emit $n$ photons, via stimulated emission, into the other ($\omega _{2},\mathbf{k}_{2}$) where \mbox{$|\mathbf{k}_{1}|=|\mathbf{k}_{2}|=k=2\pi /\lambda $}, and the detuning is \mbox{$\delta =\omega _{1}-\omega _{2}$}. This results in the transfer of momentum \mbox{$\hbar|\mathbf{q}|=nP_{recoil}\equiv 2n\hbar k\sin(\theta /2)$} and energy $\hbar n\delta $, where \mbox{$\mathbf{q}$ $=\mathbf{k}_{1}-\mathbf{k}_{2}$}, and $\theta $ is the angle between the two beams forming the grating. This process will be resonant for non-interacting stationary atoms if the final kinetic energy of the atom is equal to the energy difference between the Bragg beams i.e.~\mbox{$\hbar n$$\delta =(nP_{recoil})^{2}/2m$}.

When calculating the resonance condition for atoms which are initially moving in the lab frame an additional term arises, which can be attributed to the Doppler effect, and which we will call the Doppler term $\hbar\mathbf{P}\cdot\mathbf{q}/m$ (where $\mathbf{P}$ is the condensate momentum). In the micro-motion case we have \mbox{$\mathbf{P}=p(t)\mathbf{\hat{i}}$} giving a Doppler term which results in a new time-dependent Bragg detuning condition. For first order Bragg scattering ($n=1$) this condition is

\begin{equation}
\hbar \delta =\frac{P_{recoil}^{2}}{2m}-\frac{\hbar \left\vert \mathbf{q}%
\right\vert \mu B_{q}^{\prime }}{m\omega _{T}}\sin (\omega _{T}t).
\label{timedepdelta}
\end{equation}%
\noindent Note that when $\delta$ is greater than $P_{recoil}^{2}/2m\hbar$ the resonance condition can only be satisfied for \mbox{$\sin (\omega _{T}t)<0$}, so that Bragg scattering is resonant only for some instant during half of the TOP cycle. Similarly, if $\delta<P_{recoil}^{2}/2m\hbar$ resonance only occurs if \mbox{$\sin (\omega _{T}t)>0$}. Therefore, if the Bragg pulse length, $T_{B}$, is not an integer multiple of the TOP period the Bragg spectrum will be asymmetric. We avoid this problem by having $T_{B}$ equal to eight TOP rotations.

Alternatively, we can picture the Bragg scattering process in the rest frame
of the condensate and think of the micro-motion as modifying the optical
potential. In the absence of micro-motion the effective optical potential from two counter-propagating, pulsed, laser beams is given by \cite{mfbragg(Rabi)}

\begin{equation}
V_{opt}(x,t)=\hbar V(t)\cos (|\mathbf{q}|x-\delta t),  \label{grating}
\end{equation}%
\noindent where the amplitude $\hbar V(t)$ is twice the AC stark shift at the grating maximum. In this case, the potential has a spectrum
centered at the frequency difference between the two Bragg beams, with an 
\emph{rms} half-width \mbox{$\Delta \nu _{T_{B}}\simeq 1/(2T_{B})$}. 

The effect of the micro-motion can be incorporated into the optical potential by transforming to the stationary frame of the condensate. The resulting time-dependent frequency shift of the Bragg resonance condition in the laboratory frame produces a frequency modulated (FM) optical potential in the condensates rest frame, as pointed out by Cristiani \emph{et al.} \cite{1Doptlat}. The modified potential becomes

\begin{equation}
\bar{V}_{opt}(x,t)=\hbar V(t)\cos (|\mathbf{q}|x-\delta t-|\mathbf{q}|A\cos
(\omega _{T}t)),  \label{fmgrating}
\end{equation}%
\noindent where the bar denotes the COM frame of the condensate and \mbox{$A=\mu
B_{q}^{\prime }/m\omega _{T}^{2}$}. This frequency modulation generates
side-bands in the frequency spectrum of the optical potential, yielding
multiple first-order Bragg resonances.

Bragg spectroscopy involves scanning the frequency difference $\delta$
through resonance and measuring the out-coupled fraction. To resolve
features in the condensate's Bragg spectrum requires the frequency
width of the Bragg light to be narrower than the spectral width of the
feature. Mechanisms that contribute to the condensate momentum width \cite{MITspec,Braggspec} are Doppler broadening \mbox{$\Delta \nu_{D}=\sqrt{(21/8)}(P_{recoil}/2\pi mR)$}, where $R$ is the condensate
radius, and mean-field (collisional) broadening \mbox{$\Delta \nu_{n}=\sqrt{%
(8/147)}(\mu_{max}/h)$}, where $\mu_{max}$ is the chemical potential at the
peak density. In addition to these effects and the spectral broadening
associated with finite pulse time, there is also a power broadening
contribution \mbox{$\Delta \nu_{p}\simeq kV/\pi |\mathbf{q}|$} \cite{mfbragg(Rabi)}.

\section{\label{expt}Experimental Details}

Our apparatus for producing Bose condensates is described in \cite%
{martin}, but with some minor modifications. Briefly, we use a double
magneto-optical trap (MOT) to cool and trap $^{87}$Rb atoms, which is driven
by injection-seeded diode lasers and with continuous loading of the
low-pressure MOT. The laser cooled sample is transferred to a magnetic TOP
trap and evaporatively cooled to produce a condensate containing
approximately \mbox{$2\times 10^{4}$~atoms} in the \mbox{$\left\vert
F=2,m_{F}=2\right\rangle $} hyperfine ground state.

Following condensate formation, the magnetic trap was adiabatically relaxed
(by reducing the quadrupole field gradient and increasing the bias field
over 200~ms) to trapping frequencies of \mbox{$\omega _{radial}=\omega _{z}/\sqrt{8}=2\pi \times 18$~Hz}. For our bias field rotation
frequency \mbox{$\omega _{T}=2\pi \times 2.78$~kHz}, the corresponding tangential micro-motion velocity is 3.3~mm~s$^{-1}$ (compared to the
two-photon recoil velocity of 11.8~mm~s$^{-1}$). Although the amplitude of
the motion is only 190~nm, the high relative velocity means
that the effect of the micro-motion on the Bragg spectrum is substantial.

Counter-propagating Bragg beams were generated using a Coherent MBR
Ti:sapphire laser, the output of which was split and passed through two acousto-optic modulators (AOMs) to provide individual frequency control. We used a detuning \mbox{$\Delta \simeq $ $2\pi \times 4.49$~GHz} above the \mbox{$^{5}S_{1/2} $ $F=2$} to \mbox{$^{5}P_{3/2}$ $F^{\prime }=3$} transition, giving a spontaneous scattering rate of approximately $5$~s$^{-1}$ (i.e.~negligible for our purposes). Both beams were aligned in the plane of the rotating bias field. A pulse length of \mbox{$T_{B}=2.88$~ms} was chosen for a spectral width $\Delta\nu _{T_{B}}$ comparable to the width of an equivalent stationary condensate with our parameters. The pulse length is also shorter than a quarter trap period so that the Bragg scattered
fraction becomes well separated from the unscattered component in time-of-flight. The intensity was set between 0.26(5) to 0.6(1)~mW~cm$^{-2}$. To minimize the effect of timing differences between consecutive measurements we synchronized the Bragg pulse with the rotation of the bias field. At the end of the pulse, the condensate was
released from the trap and imaged by resonant absorption after $9$~ms of
free expansion. Bragg spectroscopy was performed by measuring the
first-order scattered fraction as a function of the frequency difference
between the two Bragg beams.

\section{\label{results}Results and Discussion}

\begin{figure}
\includegraphics[width=0.9\linewidth]{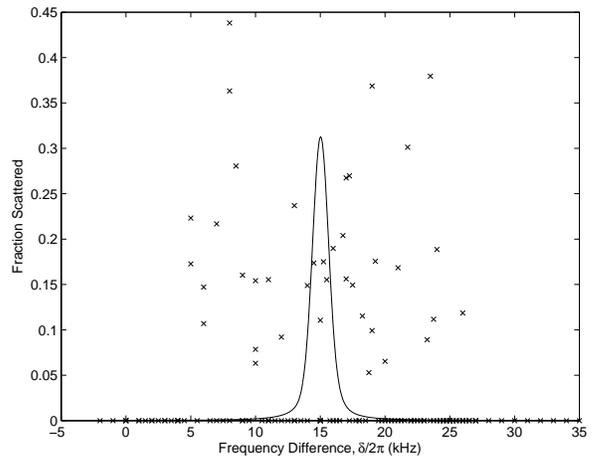}
\caption{Bragg scattered data for a Bose condensate accelerating due to
micro-motion in a TOP trap (crosses), together with a numerical simulation
using the linear 1D Gross-Pitaevskii equation assuming
no micro-motion (solid line). Results are for \mbox{$B_{q}^{\prime }$ = 90~G~cm$^{-1}$} (corresponding to a maximum COM velocity of 3.3~mm~s$^{-1}$) and an intensity of 0.26(5)~mW~cm$^{-2}$ (corresponding to a power
broadening contribution of 0.22(4)~kHz).}
\label{subplot+tof}
\end{figure}

For a stationary Bose condensate confined in a static harmonic potential
with our experimental parameters, the Bragg spectrum has a simple form
(essentially Gaussian) with a narrow width. However, the presence of
micro-motion introduces substantial broadening and complex structure. In
Fig.~\ref{subplot+tof} we show the Bragg scattered data obtained for a
condensate undergoing centripetal acceleration in the plane of the Bragg
beams. Also shown is a theoretical spectrum for our condensate parameters, ignoring mean-field effects (which are small in this case) and assuming no micro-motion. The 
\emph{rms} half-width of the theoretical curve is 0.56~kHz, which is
consistent with our measurements made in time-of-flight where micro-motion is
absent.

\begin{figure}[tbp]
 \includegraphics[width=0.9\linewidth]{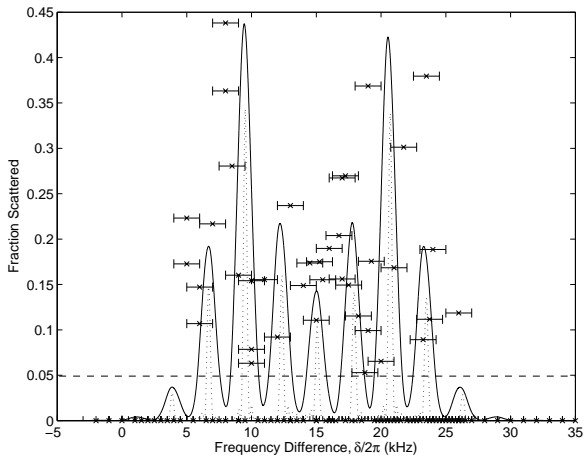}
 \caption{\label{datamodel}The Bragg scattered data from Fig.~\protect\ref{subplot+tof} (crosses),
together with the spectrum of the Bragg light (dotted line) and the
corresponding numerical simulation (solid line) using the linear 1D
Gross-Pitaevskii equation. The relative heights of the peaks in the FM
spectrum are determined by the modulation index and the width of each peak
is determined by the finite pulse length $T_B$. Each data point has an uncertainty in the scattered fraction of $\pm~0.1$. The dashed line indicates the minimum detectable fraction for this data and below this limit we plot the data as having zero scattered fraction.}

\end{figure}

To explain the complex structure in the experimental spectrum, we now
present results of the simple FM picture described in the previous
section. 
We obtain the frequency components present in the Bragg light by performing a
Fourier transform on the modified optical potential given by Eq.~(\ref
{fmgrating}). The result is a spectrum for the Bragg light which has a
carrier at the frequency difference between the Bragg beams ($\delta$), and side-bands (spaced by $\omega _{T}$) with relative amplitudes
determined by Bessel functions of the FM spectrum \cite{fm}. In this case the
modulation index, upon which the Bessel functions are strictly dependent, is
just the maximum Doppler shift divided by the modulation frequency and can
therefore be adjusted by simply changing the quadrupole field gradient. In
Fig.~\ref{datamodel} the experimental data from Fig.~\ref{subplot+tof} is
overlaid with the corresponding frequency spectrum of the modified optical
potential $\bar{V}_{opt}$ (scaled by an arbitrary factor). Since there are a number of side-bands with significant amplitude, when we sweep the carrier frequency ($\delta$) to generate the Bragg spectrum we
bring successive side-bands into resonance with the condensate. The exact
form of the Bragg spectrum is then a combination of the complex spectral
character of the light and the spectral response of the stationary
condensate. When the Bragg spectrum of the stationary
condensate is narrow compared to the bias field rotation frequency (as in
our case), the Bragg spectrum with micro-motion has peaks at frequencies
corresponding to the locations of the side-bands. To within the spectral
uncertainty of our Bragg experiments, approximately $\pm 1$~kHz, this
behavior is shown in the data of Fig.~\ref{datamodel}. Although we cannot resolve the
predicted individual spectral peaks, the overall range of frequencies for
which we get Bragg scattering matches the range over which there is
significant sideband intensity.  Our spectral resolution is limited by a
number of physical processes, including the mechanical stability of the
optics and the electrical stability of the currents driving the magnetic
trap. From a heterodyne measurement using the two Bragg beams we
determined that the uncertainty due to the mechanical
stability of the optics (which Doppler shifts the light) is approximately $
\pm 0.4$~kHz, whereas the contribution to the uncertainty from current
stability is approximately $\pm 0.6$~kHz. The main effect of current instability
is to put noise on the quadrupole gradient, which in turn puts noise on the
time-varying COM momentum, as can be seen from Eq.~(\ref{oscilmomentum}). The
uncertainties in our measurements of the fraction of scattered atoms were
determined by making repetitive measurements with the same parameters, and can vary across the spectrum. The significant contributions arise from shot-to-shot condensate variation and the minimum scattered fraction we can measure with our imaging system.

At low intensity the simple FM picture predicts some important features such as the frequencies and relative heights of the peaks, but the widths of each peak only include a contribution from the finite pulse time. To obtain
a more accurate prediction of the experimental results we include the
dynamic term of the TOP trap in a collisionless one-dimensional
Gross-Pitaevskii equation simulation, using a one-dimensional version
of Eq.~(\ref{instanpot}) as our magnetic potential. Results of this simulation
are shown as the solid line in Fig.~\ref{datamodel}, confirming aspects of
the FM picture and providing better agreement with the data.
\begin{figure}[tbp]
\includegraphics[width=0.9\linewidth]{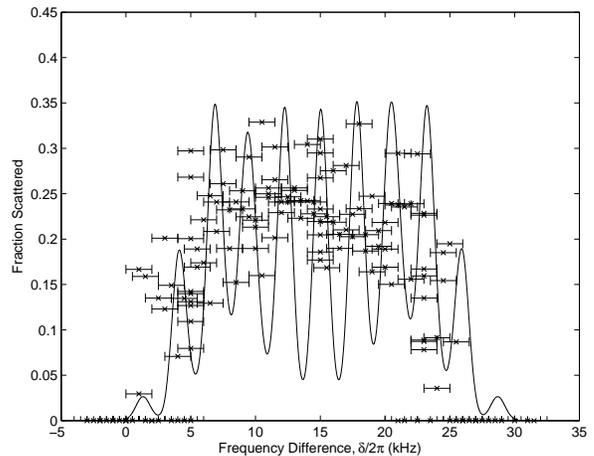}
\caption{\label{gpe+data}Power-broadened Bragg scattered data (crosses) and the corresponding 1D
linear GPE simulation (solid line). Experimental parameters are \mbox{$B^{\prime}_{q}$ = 90~G~cm$^{-1}$} and a intensity of 0.6(1)~mW~cm$^{-2}$ (corresponding to a power broadened contribution of 0.5(1)~kHz). Each data point has an uncertainty in the scattered fraction of between $\pm~0.05$ to $\pm~0.09$ depending on $\delta/2\pi$.}
\end{figure}

As an additional test of our interpretation we now consider a power
broadened case (i.e.~a higher value for $V(t)$). This alleviates the need to resolve very narrow spectral features and gives larger (more easily
measured) Bragg scattered fractions near the central region of the spectrum. Fig.~\ref{gpe+data} shows a power
broadened Bragg spectrum together with the corresponding numerical
simulation. The contribution to the condensate width from power broadening is $0.5(1)$~kHz, compared to $0.26$~kHz from Doppler broadening (the dominant condensate
contribution). Within the limits of our experimental uncertainty, the
agreement between the data and the simulation is reasonable. The overall
width of the spectrum is well predicted, and the variation in the Bragg scattered
fraction is reduced, as expected. The small difference between the center
frequencies of the GPE simulations and experimental data spectra is likely to be due to our linear one-dimensional modeling of a nonlinear three-dimensional experiment \cite{Morgan,mfbragg(Rabi),Braggspec}. The slight asymmetry in
the numerical spectrum (which we cannot resolve in the experimental data)
becomes apparent when transforming to the condensate rest frame through a
damped drift term \cite{kath}. Additional asymmetry can occur if
the Bragg pulse length does not match an integer multiple of the TOP trap
rotation period, but this is not the case for the results presented here.

A process not included in our linear GPE simulations, but which can
significantly alter the Bragg spectrum \cite{mfbragg(Rabi)}, is the mean field collisional interaction in the condensate. A detailed theoretical treatment is currently being prepared \cite{kath}.

\section{\label{conclu}Conclusions}

We have presented experimental results and a simple theoretical analysis
of Bragg spectroscopy performed on an accelerating Bose-Einstein
condensate.  This work extends Bragg scattering experiments beyond those
involving Bose condensates with a stationary (or constant velocity)
center of mass.  Using condensate micro-motion in a magnetic TOP trap,
we have shown that for circular center of mass motion the Bragg spectrum
is modified significantly from a simple momentum interpretation.  An
analysis which treats the angular acceleration as equivalent to
frequency modulating the optical grating provides a useful
interpretation for the complex structure we observe in the Bragg spectra. Numerical simulations using the one-dimensional linear Gross-Pitaevskii equation extend the analysis to include other broadening mechanisms. Although we are not able to resolve the narrow features predicted by our models, the overall spectral width and scattered fractions are in quantitative agreement with the theoretical results, confirming our simple physical interpretation. However, a more comprehensive theoretical
description of Bragg spectroscopy with accelerating Bose condensates is
needed to characterize the effects of collisions and collective
excitations in higher dimensions.

\begin{acknowledgments}
The authors thank K.~J.~Challis for assistance with computational and theoretical work, G.~Duffy for help with data acquisition, and R.~J.~Ballagh for useful comments and suggestions. We gratefully
acknowledge the financial support of the Royal Society of New Zealand
Marsden Fund (contract UOO508). 

\end{acknowledgments}

\end{document}